\documentclass[fleqn,usenatbib]{mnras}
\usepackage{natbib}
\usepackage{graphicx}
\usepackage{color}
\usepackage[dvipsnames]{xcolor}
\usepackage{amsmath}
\usepackage{amssymb}
\usepackage[T1]{fontenc}
\usepackage{grffile}
\usepackage[normalem]{ulem}
\newcommand{\higpus}{\texttt{HiGPUs~}}

\title[Short title, max. 45 characters]{MNRAS \LaTeXe\ template -- Formation of a NSC in NGC 4654}

\title[Formation of a NSC in NGC 4654]{Are we observing a NSC in course of formation in the NGC\,4654 galaxy?}

\author[R. Schiavi et al.]{
R. Schiavi,$^{1}$\thanks{E-mail: riccardo.schiavi@uniroma1.it}
R. Capuzzo-Dolcetta,$^{1}$
I. Y. Georgiev,$^{2}$ 
M. Arca-Sedda$^{3}$
\newauthor
\hspace*{.2cm}and A. Mastrobuono-Battisti$^{2,4}$\\
\\
$^{1}$Dipartimento di Fisica, Universit\`a degli Studi di Roma ``La Sapienza'', P.le Aldo Moro, 2, I-00185 Rome, Italy \\
$^{2}$Max Planck Institute for Astronomy, K\"onigstuhl 17, D-69117 Heidelberg, Germany\\
$^{3}$Astronomisches Rechen Intitut, Zentrum f\"ur Astronomie der Universit\"at Heidelberg,\\ M\"onchhofstrasse 12-14, D-69120 Heidelberg, Germany\\
$^{4}$Department of Astronomy and Theoretical Physics, Lund Observatory, Box 43, SE--221 00, Lund, Sweden
}

\begin{document}
\date{\today}
\pubyear{2015}
\label{firstpage}

\maketitle

\begin{abstract}
 We use direct $N$-body simulations to explore  some possible scenarios for the future evolution of two massive clusters observed toward the center of NGC\,4654, a spiral galaxy with mass similar to that of the Milky Way. Using archival HST data, we obtain the photometric masses of the two clusters, $M=3\times 10^5$ M$_\odot$ and $M=1.7\times 10^6$ M$_\odot$, their half-light radii, $R_{\rm eff}\sim4$ pc and $R_{\rm eff} \sim 6$ pc, and their projected distances from the photometric center of the galaxy (both $<22$ pc). The knowledge of the structure and separation of these two clusters ($\sim 24$ pc) provides a unique view for studying the dynamics of a galactic central zone hosting massive clusters. Varying some of the unknown clusters orbital parameters, we carry out several $N$-body simulations showing that the future evolution of these clusters will inevitably result in their merger. We find that, mainly depending on the shape of their relative orbit, they will merge into the galactic center in less than 82 Myr. In addition to the tidal interaction, a proper consideration of the dynamical friction braking would shorten the merging times up to few Myr. We also investigate the possibility to form a massive NSC in the center of the galaxy by this process. Our analysis suggests that for low eccentricity orbits, and relatively long merger times, the final merged cluster is spherical in shape, with an effective radius of few parsecs and a mass within the effective radius of the order of $10^5\,\mathrm{M_{\odot}}$. Because the central density of such a cluster is higher than that of the host galaxy, it is likely that this merger remnant could be the likely embryo of a future NSC.

\end{abstract}



\section{Introduction}\label{sect:intro}

\begin{figure}
    \centering
    \includegraphics[width=\columnwidth]{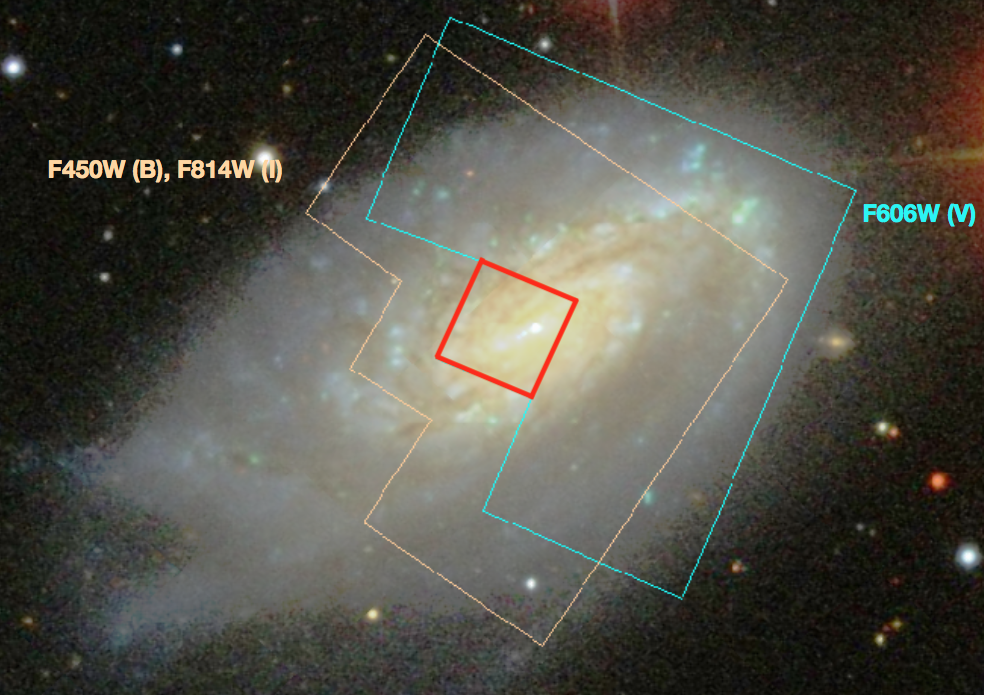}
    \caption{An SDSS colour image of NGC\,4654 overlaid with the HST\-/\-WFPC2 fields of the two available filters. The analysis in this paper is based on the $F606W$ pointing {\bf(outlined in cyan}) with the nucleus centered on the high spatial resolution WFPC2/PC-chip shown with red rectangle ($\sim2.2\times2.2$\,kpc). A further zoom in the nuclear region ($100\times100$\,pc) around the two clusters is shown in Figure\,\ref{fig:obs_2D}.}
    \label{fig:N4654_SDSS_WFPC2_FoV}
\end{figure}

Nuclear star clusters (NSCs) are dense and massive clusters observed  with high frequency ($>80\%$) at the centre of galaxies with stellar masses $10^9-10^{11}$ M$_\odot$ \cite[e.g.][]{Boeker02,Boeker04,Cote06,Turner12,GB14,denBrok14,Baldassare14,Sanchez-Janssen19,Pechetti20}.
These extreme environments often harbor a central supermassive black hole (SMBH, as in the case of our Galaxy), and represent the most dense stellar systems in the Universe.  We refer to the recent and complete review by \cite{Neumayer_2020} for more details about NSCs.
There are two main formation channels that are thought to compete in NSC formation: {\it in-situ} formation via fragmentation of gaseous clouds in the galactic centre \citep[e.g.][]{loo1982}, or via orbital segregation and merger of massive star clusters that migrate toward the galactic centre via dynamical friction \citep{Tre1975,RCD1993}. 
The latter formation channel, named {\it dry-merger} scenario, has been widely explored theoretically and numerically \citep{Mio2008a,Mio2008b}. For instance, high-resolution $N$-body models suggested that the Milky Way (MW) NSC might have formed through this mechanism \citep{ant2012,Tsatsi17,Arcasedda20}, which might explain both the structure and kinematics of the Galactic NSC. The {\it dry-merger} scenario provides also a successful explanation for the potential formation of NSCs in young galaxies and for the seemingly absence of nucleated regions in small dwarf  and massive ellipticals \citep{AS2014,Arcasedda17}. In fact, several semi-analytic models have shown that the {\it dry-merger} scenario leads to correlations between the NSC and the host galaxy properties pretty similar to the observed one \citep{Ant2013,Gnedin14,AS2014,CDTosta2017}. 
Nonetheless, several features of NSCs seem hard to explain as the result of star cluster merging events. For instance, NSCs exhibit a complex star formation history  that seems to suggest the occurrence of several episodic star formation events over the entire course of their lifetime \citep{Neumayer_2020}. Such a feature can be also easily explained with {\it in-situ} formation \citep{Boeker_2004}, thus suggesting that the formation and evolution of NSCs is likely the result of both scenarios operating in concert. 
Here we present our numerical approach to test this scenario in one of the clearest examples of two massive star clusters caught in the process of merging in the nucleus of the nearby MW-like spiral galaxy NGC\,4654 shown in Figure\,\ref{fig:N4654_SDSS_WFPC2_FoV} and a zoom in on its nucleus in Figure\,\ref{fig:obs_2D}. Their projected separations, photometric mass and assumption for the local velocity field of a MW-like galaxy suggests that they should be on a short, few tens of Myr, collision course before they completely merge \citep{GB14}. However, this pure analytical expectation needs to be tested in order to gain a 
deeper knowledge on 1) what will be the merging time-scales given the cluster current observational properties; 2) how the physical and observational properties of the final product depend on the merger dynamics and how such properties compare to those of current NSCs in galaxies of similar mass and type as NGC\,4654; 3) in the hypothesis that the two clusters contain different stellar populations, what are the expected distributions and fractions of the stars coming from the two progenitors in the merger product.

\section{Image Data and Galaxy Modeling}\label{sect:Gal_mod}

\begin{figure}
\centering
\includegraphics[width=0.9\columnwidth]{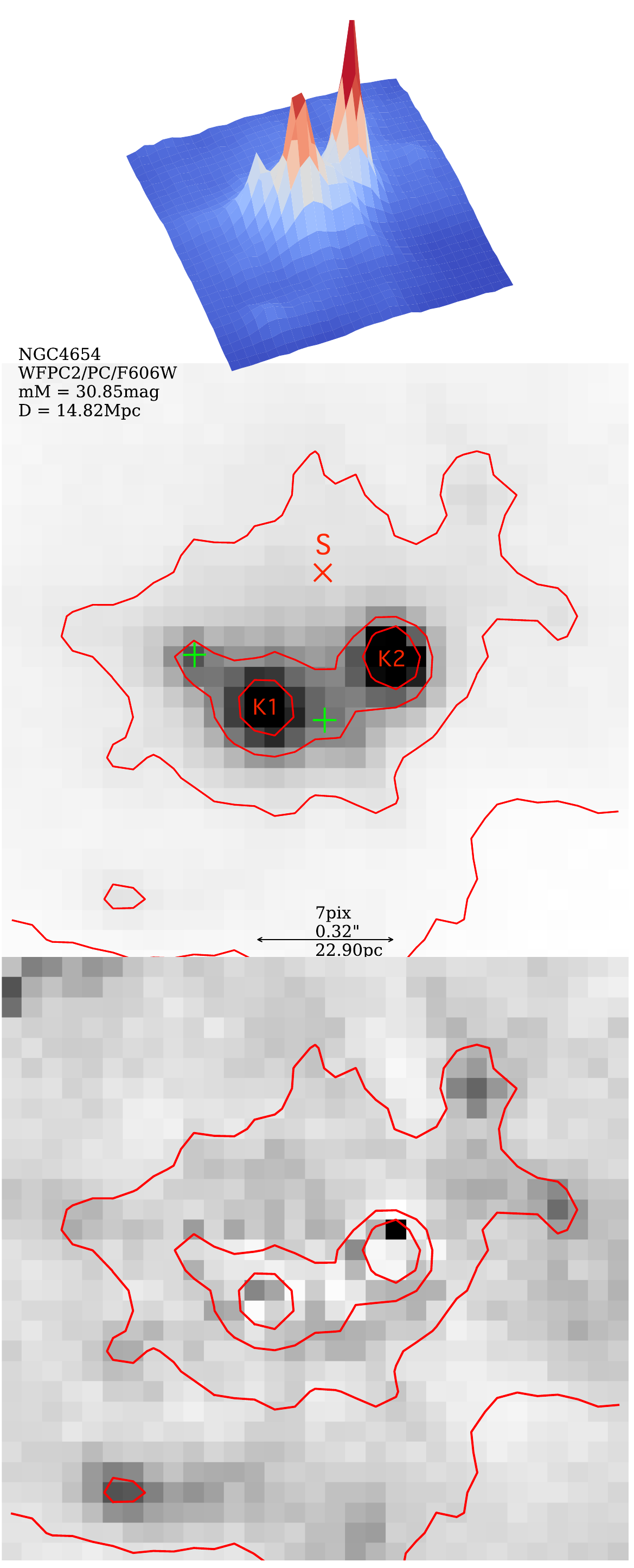}
\caption{The central $\sim\!100\!\times\!100$\,pc around the two clusters in the nucleus of NGC\,4654. The top panel illustrates the relative intensity of the data. The middle panel shows the two main clusters, K1 and K2, together with the other two fainter sources K3 and K4 (green crosses). The red cross with "S" label indicates the photometric the center of the S\'ersic profile. The bottom panel is the residual image from the best fit model (details in \S\,\ref{sect:Gal_mod}). Iso-intensity contours are shown at 3,\,10,\,30 and 50$\times$\,the image median value.}
\label{fig:obs_2D}
\end{figure}

Photometric and structural properties of the two clusters and the local galaxy background are estimated using the highest resolution (0.05\arcsec/pix\,$\sim$3.6\,pc/pix) archival HST/WFPC2/PC data, initially presented in \citet{GB14} (cf. their Fig.\,16). We use these reduced images and the specific for the detector, filter and position super-sampled {\sc TinyTim}\footnote{\url{http://www.stsci.edu/hst/instrumentation/focus-and-pointing/focus/tiny-tim-hst-psf-modeling}} \citep{Krist&Hook11} PSF to analyze the two clusters and the nuclear region of NGC\,4654 in an identical manner to \citet{Georgiev19}. Namely, we used {\sc imfit}\footnote{\url{https://www.mpe.mpg.de/~erwin/code/imfit/}}, with the PSF and a choice of analytical profiles to perform, in 2D, an iterative maximum-likelihood and $\chi^2$ minimization fitting \cite[details about {\sc imfit} in][]{Erwin15}. 
We have experimented with various profiles and their minimum required number to minimize the residuals. In Figure\,\ref{fig:obs_2D} we show the central 100 square pc of NGC\,4654, where the top panel is a surface plot, in the middle is a 2D intensity and in the bottom is the residual by subtracting the best fit model from the data.
In the middle panel of Fig.\,\ref{fig:obs_2D} we label with K1 and K2 the two clusters and with a red cross the center, S, of the galaxy fitted by a \cite{Sersic63} profile. \\ We note that after fitting for the two main clusters K1 and K2, we found small residuals (`clumps') left on the upper and lower side of the K1 cluster. To check for the significance of these residuals, we fitted two additional King components. This results in two faint components, K3 and K4, ($\sim2$\,mag fainter than K1) very hardly fitted by King profiles, due to the very low S/N values. Their location is indicated with two green crosses in the middle panel of Fig.\,\ref{fig:obs_2D}.
Because they are completely unresolved in the lower resolution WFPC2/WF3 images for $F450W$ and $F814W$, we are unable to access their color and respectively mass. However, if we conservatively assume the same stellar population content as K1, this would imply a factor of $\sim100$ less massive than K1, and even less for lower $M/L$ if they are younger. This would make them dynamically irrelevant contributors for the evolution of the system.

The final fit parameters of the relevant components (described below) and their uncertainties estimated via bootstrapping are shown in Table\,\ref{tab:fit_components}. We successfully fit the two clusters, K1 and K2, with \cite{King62} models, with core radii of 4.24 and 2.77 pc and photometric masses of $M_{K1}=1.7\times10^{6}\,\mathrm{M_{\odot}}$ and  $M_{K2}=3.2\times10^{5}\,\mathrm{M_{\odot}}$, respectively. These masses are calculated from the Galactic reddening corrected $F450W,\,F606W$ and $F814W$ model magnitudes and their color combinations, using the maximum-likelihood weighted $M/L$ from the Single Stellar Population (SSP) model grid of \citeauthor{BC03} (\citeyear{BC03}, \citeauthor{Gutkin16} \citeyear{Gutkin16} update). Here we have assumed an age of 10\,Gyr and solar metallicity for K1 and an upper age limit of 200\,Myr for K2, as can be inferred from its blue colours. The assumption of a solar metallicity is unfortunately unavoidable due to the strong age-metallicity-reddening degeneracy in the optical. Such metallicity is found to be typical for the NSC of the Milky Way and galaxies of similar mass and type \cite[e.g.][]{Rossa06,Spengler17,Feldmeier-Krause20}. The uncertainty in the masses due to the latter is of the order of 0.5\,dex. Unfortunately, the lack of either high-resolution near-infrared or optical spectroscopic data prevents us from a more precise quantification of their masses. \\
The underlying galaxy is fitted with a S\'ersic profile \citep{Sersic63} that in terms of its projected light intensity is:
\begin{equation}
    I(R)=I_e \exp\left\{-b_n\left[\left(R/R_{e}\right)^{1/n}-1\right]\right\}
    \label{sersic}
\end{equation}
where $I_e$ is the projected intensity at the effective radius $R_e$, and $b_n$ is a normalizing function of the index $n$. However, to model the galactic center we use the deprojected S\'ersic density law \citep{Merritt06}:
\begin{equation}
\label{se1}
    \rho(r)=\rho'\left(\frac{r}{R_e}\right)^{-p} \exp\left[-b\left(\frac{r}{R_e}\right)^{1/n}\right]
\end{equation}
where the parameter $\rho'$ is defined in such a way that the total mass in this formula equals that from equation \ref{sersic}. 
The $p$ parameter has been calculated with the formula by \cite{LimaNeto99}:
\begin{equation}
\label{se2}
    p=1-0.6097/n+0.05463/n^2 .
\end{equation}
Giving an effective radius $R_{e}=131.1$ pc and a n-index of 0.74, the galaxy total mass results to be $M=1.29\times10^{10}\mathrm{M_{\odot}}$.

\section{Main assumptions and uncertainties}
 Due to the lack of high spatial ($<0\farcs5\lesssim36$ pc) resolution spectroscopic data we can only assume that the photometric and kinematic center of the galaxy coincide. Such an assumption is also supported by observations for galaxies of similar mass, which also indicate that the NSC is in the photometric and kinematic centre of the galaxy, within the uncertainties of the dynamical center determination \cite[e.g.][]{Seth10, Neumayer11, Neumayer_2020}.
Thus, under the assumption that the luminosity-density center of the galaxy coincides with the kinematic center and given the relatively small uncertainties of the cluster coordinates, we conclude that they are both significantly off-center. This allows us to consider each cluster orbiting around the galactic center.
The projected distances of the two clusters, with respect of the S\'ersic center, are 21.55 and 19.8 pc respectively, giving a relative distance between them of 24.15 pc. \\
The distance to NGC\,4654 is relatively uncertain and the available measurements in the NED\footnote{http://ned.ipac.caltech.edu/} and HyperLEDA\footnote{http://leda.univ-lyon1.fr/} databases are primarily coming from the Tully-Fisher relation. The distance ranges from 13 to 19\,Mpc with a median of 14.82\,Mpc, which is the distance that we adopt here. We note, however, that due to this uncertainty range in distance, the measured effective radius and mass could vary and be biased from the median by a factor of $\sim\!0.9-1.4$ and 0.3\,dex, respectively. The adopted filter-specific foreground Galactic reddening correction for the WFPC2/F606W is of 0.063\,mag. This value is based on the \cite{Schlafly11} re-calibration of the \cite{Schlegel98} Galactic dust maps, assuming \cite{Fitzpatrick99} reddening law with $R_V=3.1$. \\

\begin{table*}
    \centering
    \begin{tabular}{c|ccccccllcc}
    \hline\hline
    Compon. & R$_{rel}$ & $F606W_0\ (V)$ & Flux frac. & $r_c$ & c & $\alpha$ & $r_{\rm eff}$ & PA & ellipticity & mass\\
            & [pc] & [mag]             & [\%]      &   [pc] & [$r_t/r_c$] & & [pc] & [deg] & (1$-b/a$) & [$\mathrm{M_{\odot}}$] \\
    (1)&(2)&(3)&(4)&(5)&(6)&(7)&(8)&(9)&(10)&(11)\\
    \hline
    King 1 & {\bf 21.55$\pm0.87$} &18.637 & 2.72 & 4.24$^{+0.14}_{-0.25}$ & 7.6$^{+1.7}_{-2.0}$ & 1.79$^{+0.27}_{-0.38}$& 6.65$^{+0.14}_{-0.25}$ & 47.1$^{18.1}_{-17.4}$ & 0.10$\pm0.04$ &1.7$\times10^6$\\ 
    King 2 &{\bf 19.80$\pm0.87$} & 19.006 & 1.93 & 2.77$^{+0.18}_{-0.14}$ & 7.4$^{+1.3}_{-0.9}$ & 1.76$^{+0.33}_{-0.31}$& 4.3$^{+0.18}_{-0.14}$ & 25.7$^{+6.8}_{-10.6}$& 0.18$\pm0.05$ &3.2$\times10^5$\\
    S\'ersic &{\bf 0.0$\pm0.87$} & 14.783 & 94.46 & \multicolumn{3}{c|}{$n=0.74\pm0.04$} & 131.1$^{+3.9}_{-4.5}$ & 135.7$^{+1.9}_{-1.8}$ & 0.36$\pm0.02$ & $1.29\times10^{10}$ \\
        \hline\hline
\end{tabular}
\\ 
    \caption{ The properties of the main components as extracted from data shown in Fig.\,\ref{fig:obs_2D}: the two clusters and the background galaxy (column 1). Column 2 gives the distance to the center of the S\'ersic profile. Column 3 gives the model magnitude corrected, only, for foreground Galactic reddening. Columns 5, 6, 7 and 8 report the King model fit core radius, concentration (tidal to core radius ratio), exponent of the generalized King fitting formula \citep{King62} and its effective radius. The $n$-index of the S\'ersic component is in a multicolumn (5-7) and its effective radius is in col. 8. Note that the S\'ersic parameters may not correctly represent the large scale galaxy parameters due to the limited fitting area on the WFPC2/PC detector.
    For all the components, the position angle (East of North) and their ellipticity are given in columns 9 and 10, while col. 11 gives their masses. All uncertainties are the posterior distributions percentiles (one-sigma) from 1000 bootstrap realizations of the fitting with {\sc imfit} \citep{Erwin15}.}
    \label{tab:fit_components}
\end{table*}

\section{Methods} \label{methods}
\begin{figure}
    \centering
    \includegraphics[width=\columnwidth]{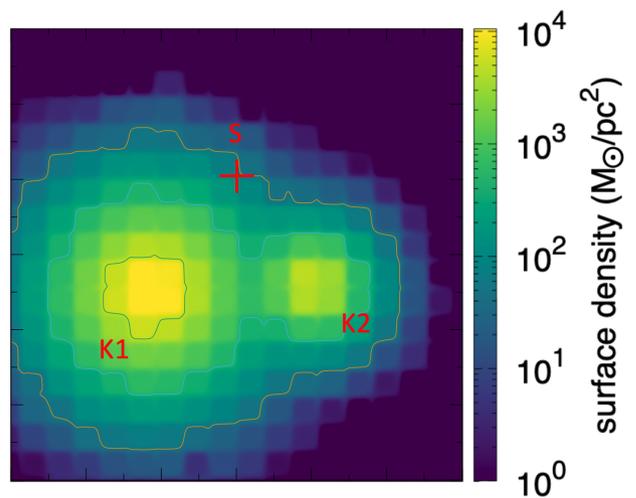}
    \caption{ The projected density map of our model in the x-y plane at the beginning of the simulations. The pixel resolution is the same as that in Fig.\,\ref{fig:obs_2D}. The red cross indicates the position of the galactic center.}
    \label{fig:init}
\end{figure}

To explore the evolution time scales of the two clusters lurking at the center of NGC\,4654 we took advantage of the \higpus code \citep{Capuzzo2013}, a full $N$-body software that exploits graphical processing units (GPUs) and implements a 6th-order Hermite scheme and block time-steps. To model the two clusters we use a total of $2.6\times10^6$ particles, thus having a mass resolution of $7.7\,\mathrm{M_\odot}$. Note that we have assumed, for the sake of simplicity, the same mass for all the stars. We adopt a softening parameter of $\epsilon=0.01$ pc to smooth out critical gravitational encounters.
We model the gravitational role of the galaxy with an analytic external potential produced by the S\'ersic profile above cited (Eqs. \ref{se1}, \ref{se2}).
The chosen frame of reference has the origin in the center of the external potential and the x-y plane coinciding with the plane of motion. For our $N$-body run, aiming to infer the time of the merger and to constrain the possible physical properties of the merger remnant, we have assumed the projected positions as actual 3D positions. Our initial distances are thus the lowest and the estimated times of merger have to be considered as lower limits. Fig.\ref{fig:init} shows the projected density map of our model. In this figure we adopted the same pixel resolution as the observations. The coordinates of the clusters with respect to the galactic center have been chosen to reproduce the same projected configuration as in Fig.\,\ref{fig:obs_2D}.

Since we do not have any observational information about the clusters' velocities, we have chosen to analyze two extreme cases. For a fixed value of the orbital energy, assumed equal to that of a circular orbit, we have taken the maximum and minimum values of the angular momentum. This leads to consider circular orbits ($e=0$) around the galactic center (C-case) and a quasi-radial fall ($e\simeq1$) towards the center (R-case). In the C-case the orbit radius is set to the current projected distance of each cluster, while in the R-case the orbits of K1 and K2 have apocenters of $27.8$ pc and $26.8$ pc respectively. \\
We note that the choice, suggested by computational convenience, of an external potential instead of a `live' $N$-body environment cannot take into account the likely effect of the dynamical friction drag of the stellar environment on the two cluster motion.  In other words, the orbital decay of the clusters in our $N$-body simulation is exclusively the result of tidal interactions, and this likely leads to a longer decay time in comparison to that expected when accounting also for dynamical friction. In any case, the uncertainty on the actual 3D positions of the clusters with respect to the galactic center is larger than that caused by neglecting the effect of dynamical friction. The main aims of this paper is to investigate whether the observed clusters can merge in the galactic nucleus in a relatively short time and to find some characteristic properties of the merger remnant. Therefore, feeling confident that the final structure of the merger remnant is largely independent of the role of dynamical friction, and considering that the merger times we obtain here are longer than the real ones, we can conclude that our results are robust for these aims.

\section{Results}\label{sect:Results}

\subsection{Merger times}\label{Sect:Merger_times}
\begin{figure}
    \centering
    \includegraphics[width=\columnwidth]{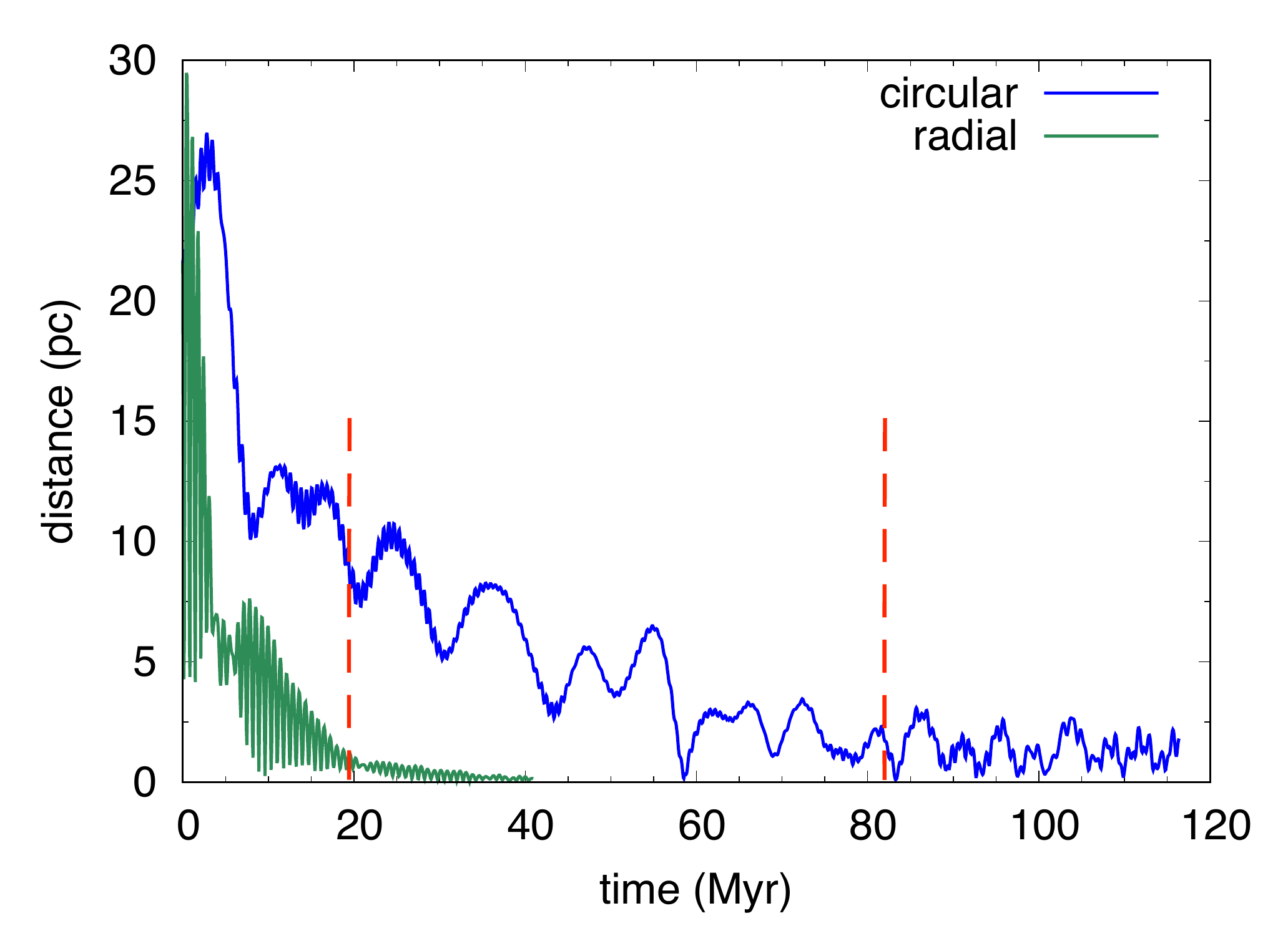}
    \caption{Distance between the centers of mass of the two clusters in the circular and radial case plotted as a function of the time. The two red dashed lines indicate the time of the merger in the R-case (left) and C-case (right).}
    \label{fig:merge}
\end{figure}

The behavior of the system composed by the two clusters K1 and K2 strongly depends on the chosen relative orbit. Fig.\,\ref{fig:merge}
shows the time evolution of the distance between the centers of mass of the two clusters in the two considered scenarios. We define the merger time as the time at which this distance becomes 0.05 times its initial value. As expected, the merger time in C-case (82.3 Myr) is about 3 times longer than in R-case (19.7 Myr).

For the sake of comparison, we note that an estimate of the dynamical friction time as based on fitting formulas in \citet{Arcasedda16} ranges for the lighter cluster K2 in the interval $6.8-641$ Myr, when 
assuming an initial circular orbit ($e=0$) and varying the initial 3D distance to the galactic center from the minimum (projected) value $19.8$ pc to a maximum assumed as $2R_e$ ($R_e=131.1$ pc being the galaxy effective radius in Eq. \ref{se1}); while, assuming an initial radial orbit ($e=1$), the range reduces to $2.2-204$ Myr. 
The uncertainty in the initial 3D distance to the center is of the order of $100$ pc. This large uncertainty justifies our choice to neglect the effect of dynamical friction in our $N$-body simulations.

\subsection{Density profiles of the merger products}\label{Sect:Results_densities}

\begin{figure*}
    \centering
    \includegraphics[width=\textwidth]{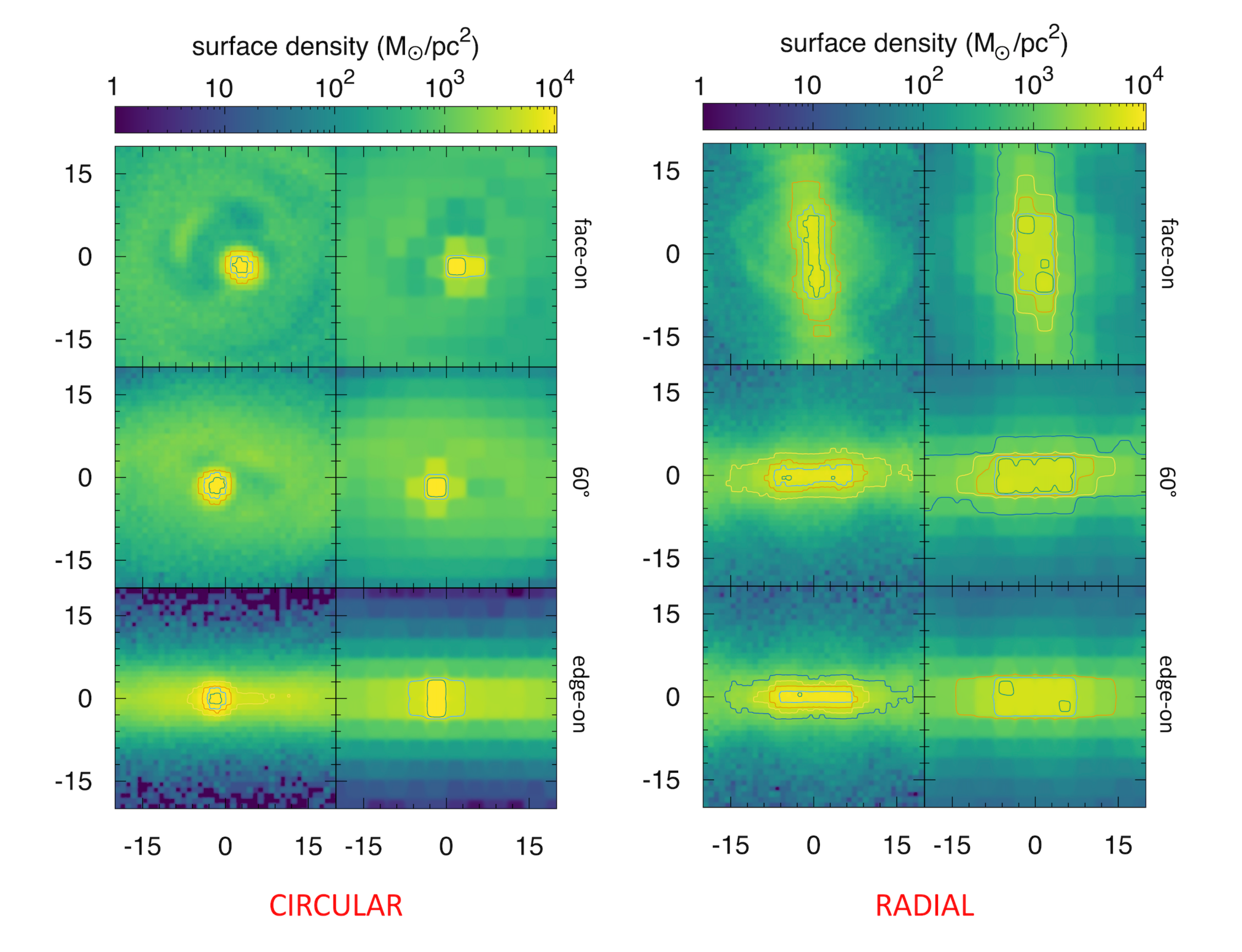}\\
    \caption{Projected density maps of the two final clusters in the C-case (left panel) and in the R-case (right panel), for two different resolutions: 1\,pc\,$\simeq0\farcs0146$\,/pix (left column) and 3.6\,pc\,$\simeq0\farcs05$\,/pix (right column). In each panel the first and the third rows show the face-on and the edge-on views respectively, while the middle row is for a line of sight at $60^{\circ}$. Axes units are pc.}
    \label{fig:final_projected}
\end{figure*}

\begin{figure*}
    \centering
    \includegraphics[width=\textwidth]{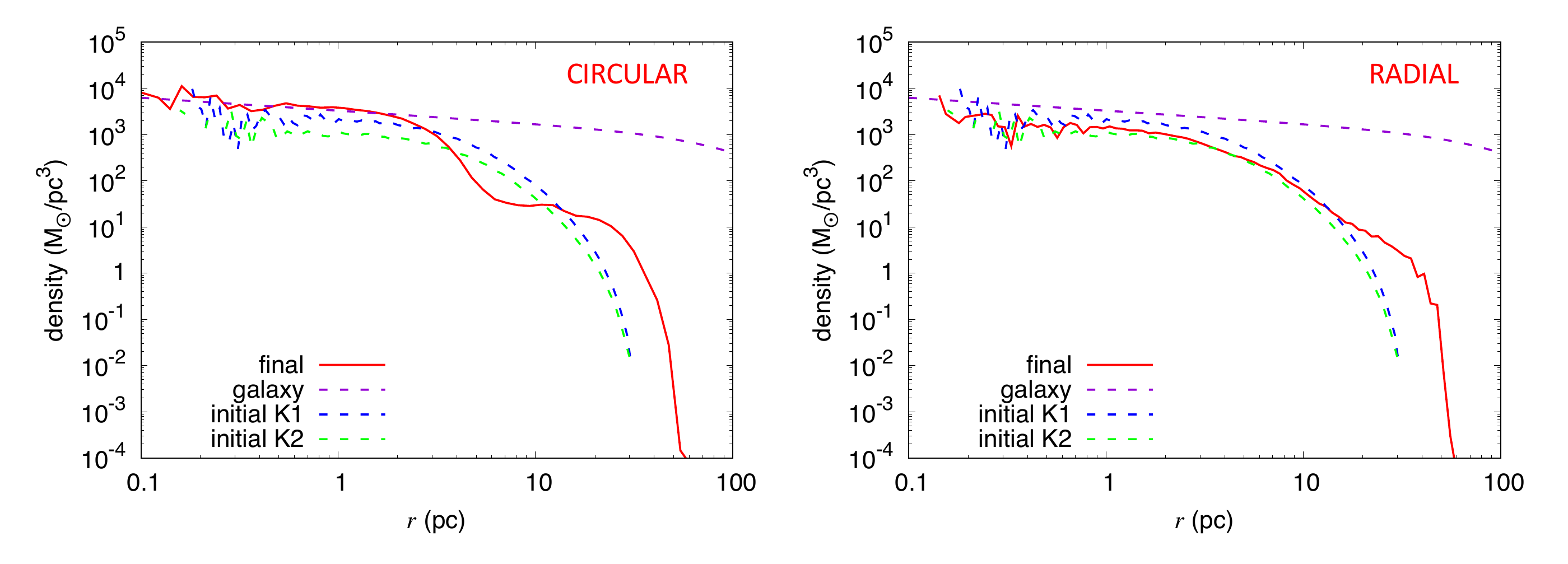}\\
    \caption{Density profiles of the two final clusters (red line) in the C-case (left panel) and in the R-case (right panel). For comparison, we also show the initial density profiles of the two clusters K1 (blue dashed line) and K2 (green dashed line), together with the S\'ersic density profile (violet line).}
    \label{fig:final_density}
\end{figure*}

\begin{figure*}
    \centering
    \includegraphics[width=\textwidth]{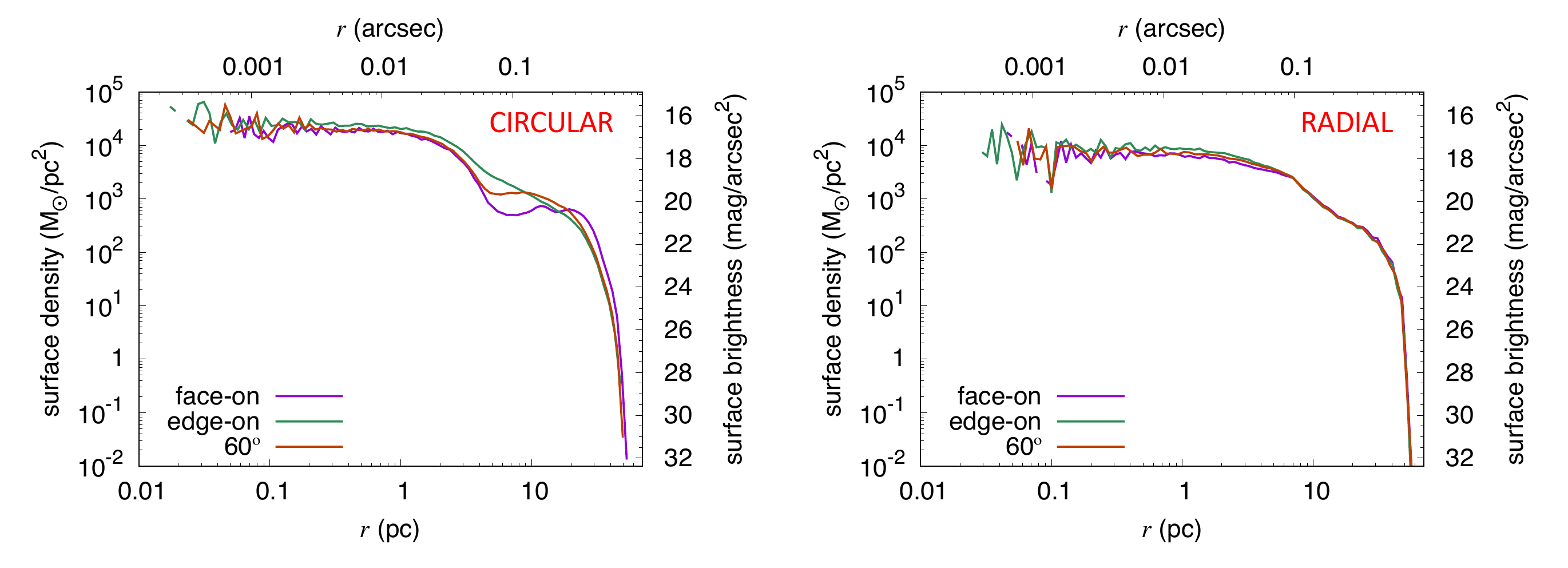}\\
    \caption{Surface density and $F606W$ surface brightness profiles of the two final clusters in the C-case (left panel) and in the R-case (right panel). Colors correspond to the three different viewing angles.}
    \label{fig:density1D}
\end{figure*}

\begin{figure}
    \centering
    \includegraphics[width=\columnwidth]{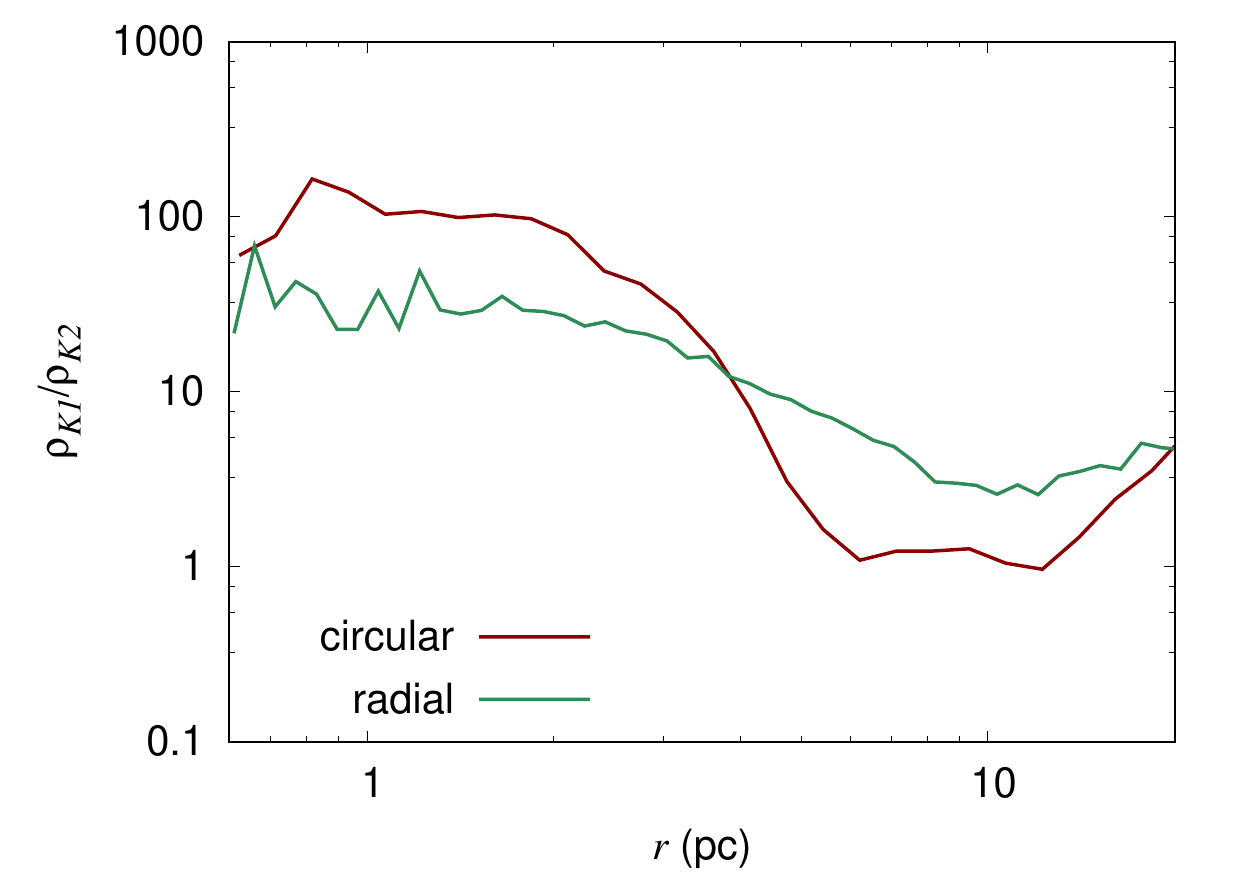}\\
    \caption{The ratio between the density of K1 and K2 stars for the circular and radial case.}
    \label{fig:ratio_stars}
\end{figure}

\begin{figure}
    \centering
    \includegraphics[width=\columnwidth]{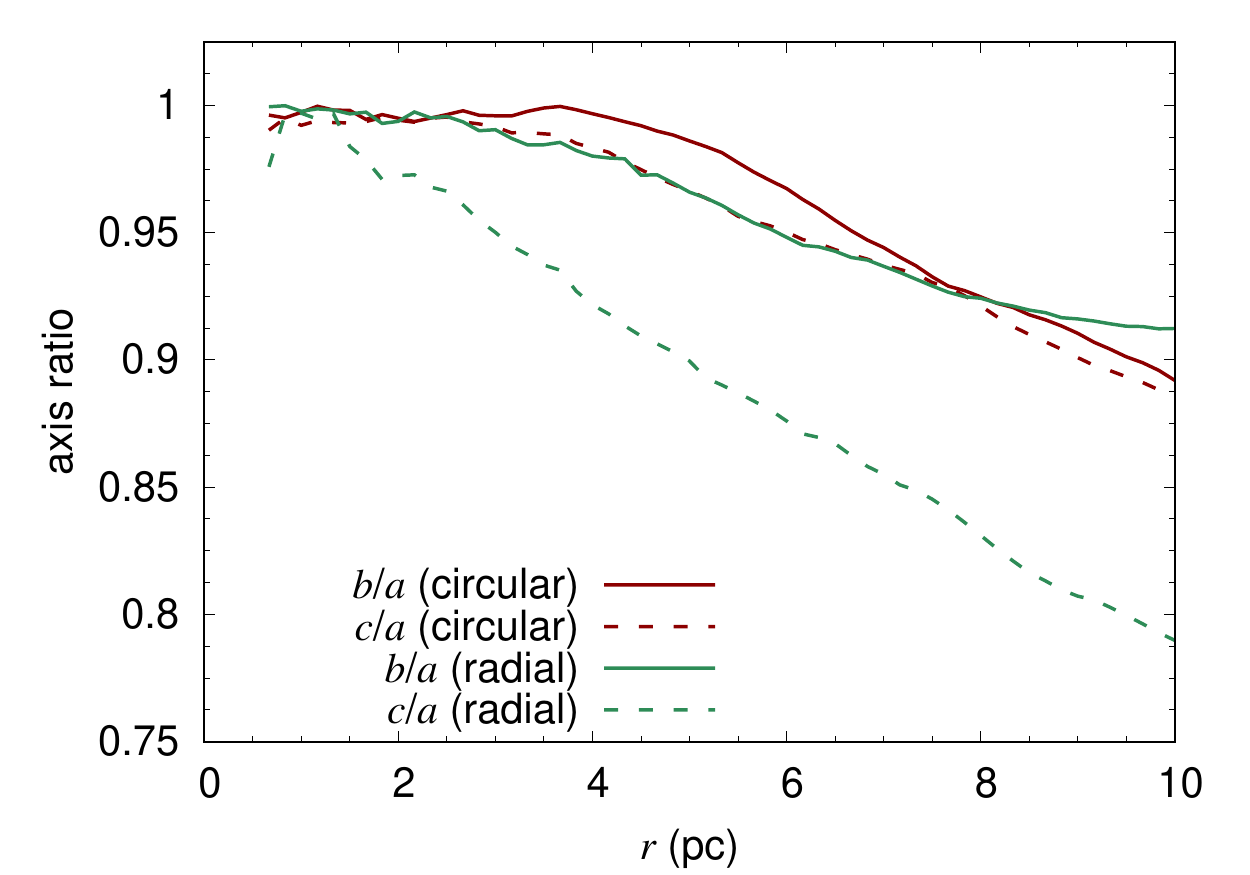}\\
    \caption{Axis ratios in the innermost 10\,pc of the merger remnant. Solid line indicates $b/a$, dashed line is for $c/a$. The colors refer to the two cases: red for circular and green for radial.}
    \label{fig:axis_ratios}
\end{figure}

The difference in the dynamics and therefore in the merger time significantly affects the shape and the density of the final cluster. In Figure\,\ref{fig:final_projected} we show the projected density maps at the end of the merger process in the two cases, using two different pixel resolutions. We have chosen three viewing angles, with respect to the z-axis, orthogonal to the plane of motion: $0^{\circ}$, corresponding to a face-on view, $60^{\circ}$ and $90^{\circ}$, that means an edge-on perspective. \\
The merger remnant in the C-case is more concentrated and looks more spherically symmetric; on the contrary, in the R-case, the final cluster appears elongated and less dense. This qualitative result is confirmed by
the density profile, in Figure\,\ref{fig:final_density}, where the density profiles of the final cluster are plotted for the two cases. For comparison we also show the observed profiles of the background galaxy and of the K1 and K2 clusters.
In the C-case the final cluster is denser and its edge is pretty well defined against the stars belonging to a more diffuse background: this edge can be identified at a distance of $\sim7$ pc. The innermost region, with a radius of $\sim2$ pc, partially emerges from the galactic profile and contains about 2 times more mass than in the R-case. On the other hand, a purely radial fall produces a more diffuse final cluster with a lower contrast on the background; we can observe, anyway, a change in the slope of the density profile at $\sim10$ pc. In this case the final cluster density remains below the galaxy profile.\\
Looking at the surface density profiles in Figure\,\ref{fig:density1D} we can notice that the pronounced contrast of the final cluster in the C-case appears mainly on the plane of motion, while on the other projection planes it is smoothed and less evident. This is mainly due to the particles lost on the longer time-scale over the orbital volume, producing a more extended low density envelope. Even though it is not spherically symmetric, the final cluster in the R-case has an almost identical average surface density profile, irrespective of the projection plane. This could be a consequence of the chaotic dynamics during the merger phases. In these plots we also show the surface brightness (SB) in the $F606W$ filter, assuming a mass-to-light ratio of $M/L=1$ and a distance to the galaxy of $D=14.82$ Mpc. Note that the underlying galaxy background light (mass) is not included to the SB-scale in Figure\,\ref{fig:density1D}.\\
It is interesting to investigate how the two species of stars, originally belonging to the clusters K1 and K2, are distributed in the merger remnant. In Fig.\,\ref{fig:ratio_stars} we show the ratio of the densities of K1 and K2 stars as a function of the distance from the cluster center. It can be seen that in the R-case the two species of stars are more evenly mixed than in the C-case, in which the central density is dominated by K1 stars. This shows that a relatively long orbital evolution, similar to that in the C-case, leads to the formation of a compact merger remnant in which stars coming from the more massive and older cluster are predominant with respect to the others. Despite in our $N$-body simulations all the particles are assumed equal, this peculiar distribution could be further explored by future studies that take into account the mass spectrum and the metallicity. This would make possiblea better comparison of our result with the NSC observations.\\
We investigate the shape of the final system by calculating the axis ratios at different distances from the center. The axis ratios are computed from the principal moments of inertia $T_{ii}$ as
\begin{equation}
    \zeta=\sqrt{T_{11}/T_{max}},\,  \eta=\sqrt{T_{22}/T_{max}},\,   \theta=\sqrt{T_{33}/T_{max}},
\end{equation}
where $T_{max}=max\left \{T_{11},T_{22},T_{33}\right \}$.
Assuming $\zeta=1$ in an appropriate coordinate system, the ratios $c/a$ and $b/a$ between the three semi-major axes are defined as the lowest and intermediate value between $(\zeta,\eta,\theta)$ respectively, being $a>b>c$. Following the same procedure described in \citet{Katz91}, we obtain the trend of the axis ratios in the innermost 10 pc. This is shown in Fig.\,\ref{fig:axis_ratios} for both the C-case and R-case.
Calculating the triaxiality parameter, defined in \citet{Antonini09} as: 
\begin{equation}
    T=\frac{a^2-b^2}{a^2-c^2},
\end{equation}
 we find that in the C-case, at 7\,pc from the center, the triaxiality parameter is $T=0.90$ and the two axis ratios are almost equal to $\sim0.94$. This means that in the C-case the final cluster has an axisymmetric shape. On the contrary, at 10\,pc from the centre, the merger remnant in the R-case has $T=0.45$ and different axis ratios: this confirms the elongated shape seen earlier.
 
On the basis of these results we can conclude that for a quasi-radial motion the process is fast and violent, and a significant part of the stars ($\sim20-30\%$) are ejected and spread through the galaxy. In the circular case, instead, the merger process is longer, but gradual; this process brings in higher amount of mass to the galactic center and leads to the formation of an almost spherical compact cluster, which could be the candidate for a future NSC. Even though we have explored only the two extreme (boundary) cases for the orbital dynamics of the observed cluster system, we find strong evidence that a relatively slow orbital decay is the fundamental condition to obtain a final compact cluster, whose physical properties are comparable with those of a NSC.

\subsection{Kinematics of the merger remnant}\label{sect:Dynamics_of_the_remnant}

To investigate the kinematics of the merger remnant in the two studied cases, we generated the radial velocity and velocity dispersion maps for the three chosen viewing angles, after subtracting the velocity components of the cluster center. Our $v_r$ and $\sigma_r$ are therefore calculated along the chosen line of sight, in a frame of reference bound to the cluster center. 

\begin{figure*}
    \centering
    \includegraphics[width=\textwidth]{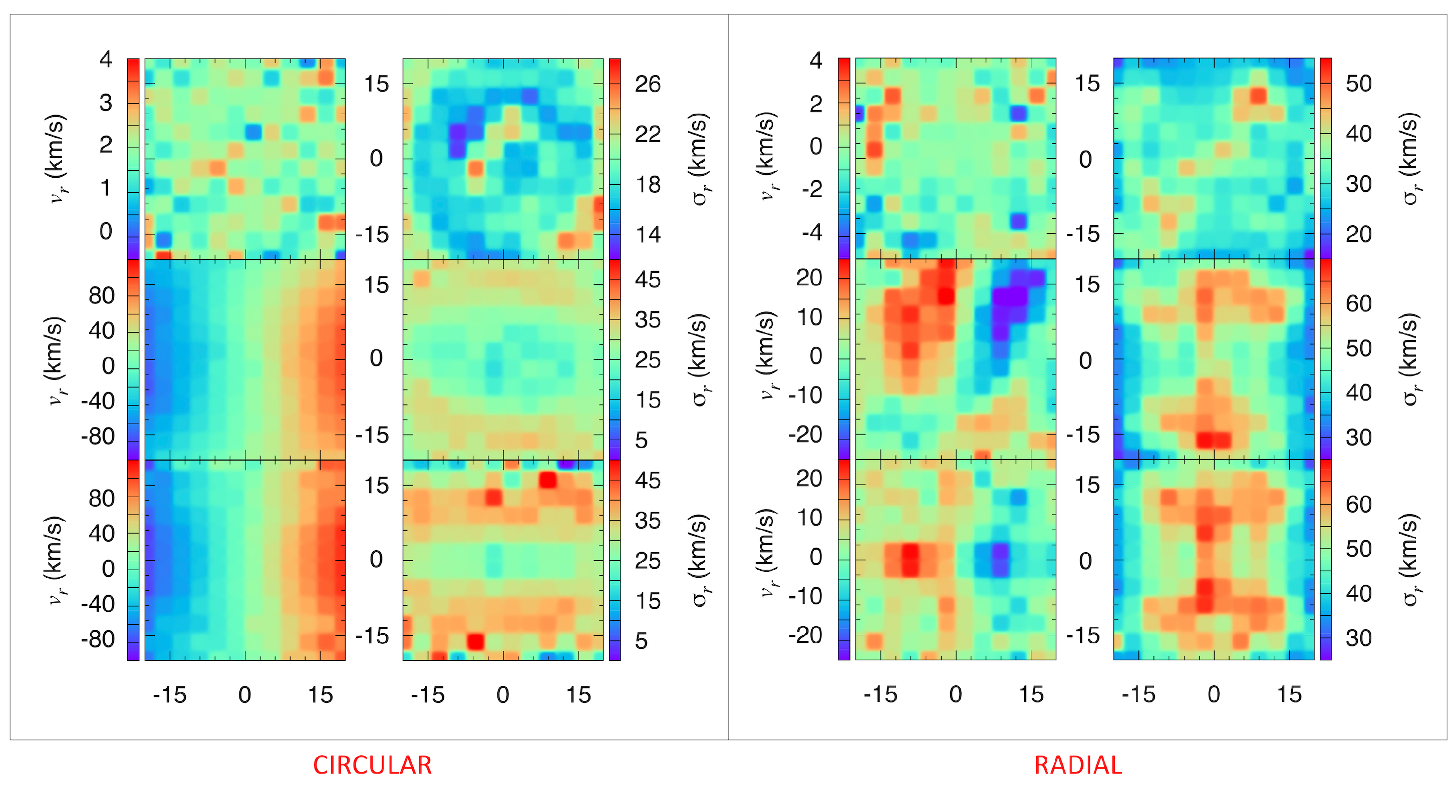}\\
    \caption{Radial velocity and velocity dispersion maps for the C-case and the R-case, for three viewing angles: face-on (first row), $60^{\circ}$ (middle row) and edge-on (third row). Note that negative velocities are for object approaching the observer, and positive velocities are for receding objects. Color bars unit is km/s, while the axes units are pc.}
    \label{fig:vel_map}
\end{figure*}

\begin{figure*}
    \centering
    \includegraphics[width=\textwidth]{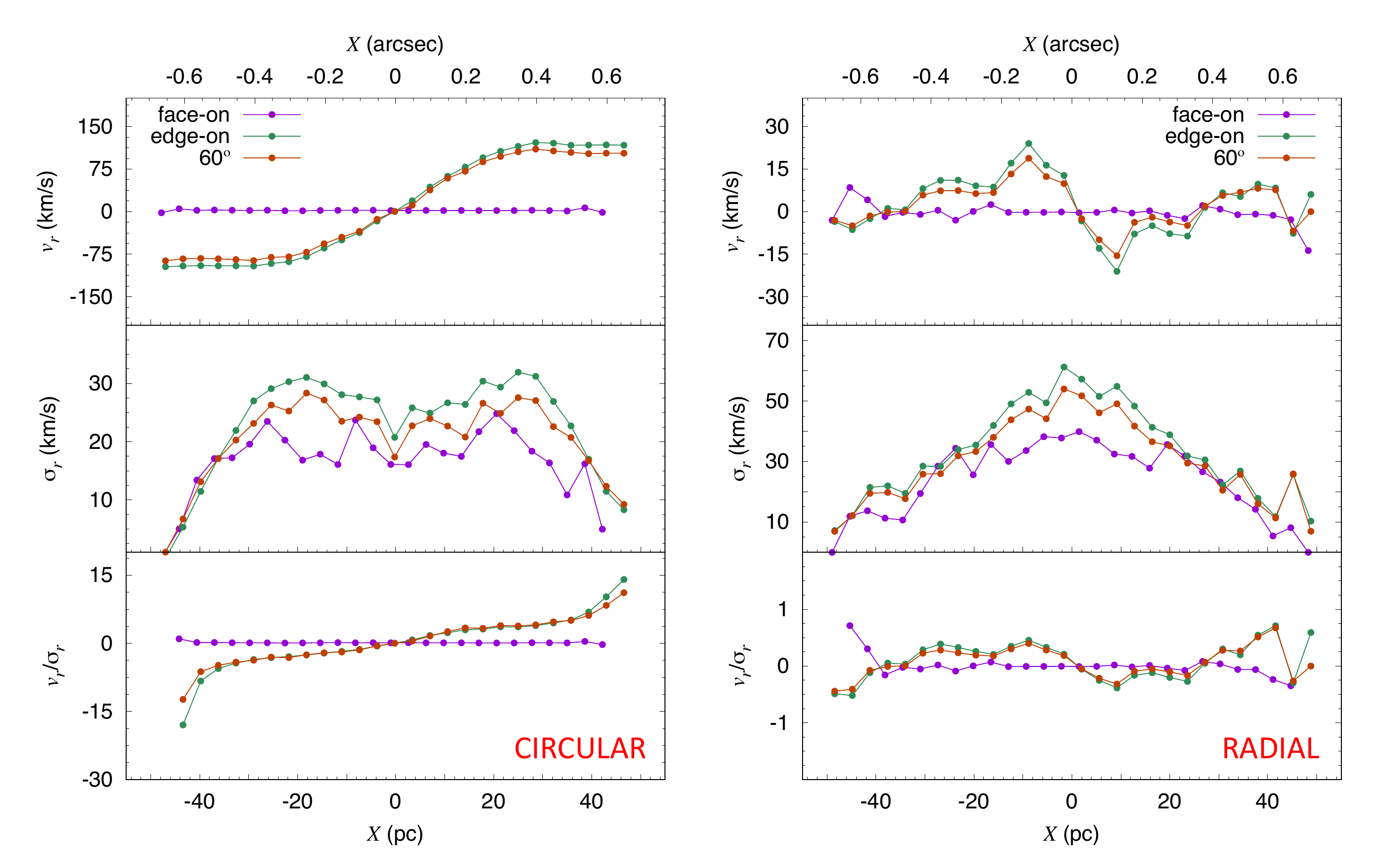}\\
    \caption{Radial velocity and velocity dispersion profiles for the C-case and the R-case, computed along three different lines of sight. Bottom panels display the ratio between $v_r$  and $\sigma_r$. Profiles are obtained using a slit width of 7.2\,pc ($\sim0\farcs1$).}
    \label{fig:vel_1D}
\end{figure*}

Looking at the radial velocity map in the C-case for the edge-on view, shown in the third row of the left panel in the Figure\,\ref{fig:vel_map}, we see that the final cluster is rotating counterclockwise on the x-y plane, with respect to the positive direction of the z-axis, in the same direction as the original clusters' orbit. The motion can be described as a rigid-body rotation in the innermost $\sim20$\,pc, as it can be inferred from the upper left panel of Figure\,\ref{fig:vel_1D}, showing the 1D radial velocity profile for the three viewing angles.

At the cluster effective radius, which is $\sim7$ pc, the rotation speed is $\sim47$ km/s, about twice the radial velocity dispersion in that region ($\sigma_r\simeq22$ km/s), while at greater distances it reaches the nearly constant values of $\sim75$ km/s.\\ In the R-case, the denser region of the final cluster, with bar-like appearance,
is slowly rotating clockwise on the x-y plane ($v_r<20$ km/s), as evident in the third row of the right panel in the Fig.\,\ref{fig:vel_map} that shows the $v_r$ map in the edge-on view. From the upper right panel in Fig.\,\ref{fig:vel_1D}, we can deduce that the rigid-body rotation of the bar is limited at the innermost $\sim10$ pc, which coincides with the cluster effective radius in this scenario, while the outer parts gradually loose any ordered motion moving far away from the center. This means that, due to the quick and violent interaction between the two clusters in this scenario, the majority of the stars are randomly moving across space. As it can be seen from the lowest right panel in Fig.\,\ref{fig:vel_1D}, showing the ratio $v_r/\sigma_r$, we can conclude that in the in the R-case the rotation is almost insignificant with respect to the random motion of the stars. \\
The rotation curves are obtained with the same method used for analysis on observational data. We have chosen a slit in the projection plane, passing through the cluster center and aligned to the intersection with the plane of the motion. The profiles shown in Fig.\,\ref{fig:vel_1D} are obtained for a slit width of 7.2\,pc, corresponding to an angular width of $0\farcs1$. Repeating the analysis for two other slit widths (3.6\,pc and 36\,pc) we found that the $v_r$ and $\sigma_r$ profiles do not change significantly.\\
We summarize the main results of the C-case and R-case in Table\,\ref{tab:results}.

\begin{table*}
    \centering
    \begin{tabular}{c|ccccccccc}
    \hline\hline
    Case & $R_{e}$ & $M_{e}$ &$\rho_{0}$ & $v_{r}$ & $\sigma_r$ &$M_{K1}/M_{e}$ &$b/a$ &$c/a$ &$T$\\
     & [pc] & [$\mathrm{M_{\odot}}$]  &  [$\mathrm{M_{\odot}/pc^3}$]  &  [km/s] &  [km/s] & & & & \\
    (1)&(2)&(3)&(4)&(5)&(6)&(7)&(8)&(9)&(10)\\
         \hline
    C-case & 7 &$1.8\times10^5$ &$1.1\times10^4$ &47 &22 &0.77 &0.944 &0.937 &0.895\\ 
    R-case &10 & $3.5\times10^5$ & $5.1\times10^3$ & 18 & 55 & 0.67 &0.912 &0.790 &0.445\\
        \hline\hline
\end{tabular}
\\ 
    \caption{Main physical and dynamical properties of the merger remnant in the two cases indicated in column (1). The effective radius, $R_{e}$, in column (2) is obtained from the inflection point in the density profiles shown in Fig.\,\ref{fig:final_density}. Column (3) gives the mass inside $R_{e}$ and Column (4) the central density. Column (5) contains the rotational velocity at the cluster effective radius, while column (6) gives the weighted mean of the radial velocity dispersion inside $R_{e}$: both are averaged for the three lines of sight. Column (7) shows the ratio between the stars that belonged to the K1 cluster and the mass enclosed in $R_{e}$. Note that here we do not take into account the mass and the density of the galaxy. Columns (8), (9), and (10) give the two axial ratios and the triaxiality parameter $T$, respectively.}
    \label{tab:results}
\end{table*}

\section{Conclusions}\label{Sect:concl}

In this paper we have presented a series of $N$-body simulations using the \higpus code \citep{Capuzzo2013} aimed at the study of the possible evolutionary path of the two massive star clusters K1 and K2 found in the central 100x100\,pc in the spiral galaxy NGC\,4654 \citep{GB14}.
Starting from initial conditions and structural parameters as deduced from the observations, and making a few simplifying assumptions on the unknown properties (see sect. \,\ref{sect:Gal_mod}) we obtained the following results.

\begin{enumerate}

\item We find a merger time interval of $19.7-82.3$\,Myr for the two clusters to merge in the galactic center, under the assumption that the true distances between the two clusters and the galactic center are their observed projected distances. The merger time interval depends on the initial orbital eccentricity of the clusters (details in \S\,\ref{Sect:Merger_times}, cf Table\,\ref{tab:results});
\item for a low eccentricity orbit, the merger process takes a longer time than for an almost radial orbit, but a higher amount of mass ($M\simeq2\times10^5\,\mathrm{M_{\odot}}$) is deposited to the innermost few parsecs ($\sim7$ pc) from the galactic center. The resulting cluster formed in this way has an almost spherical shape, a high central density ($\rho_{0}\simeq1.1\times10^4\,\mathrm{M_{\odot}/pc^3}$) and fast rotation ($>45$ km/s), as compared to the velocity dispersion ($\sigma_r\simeq22$ km/s), i.e.  $v_r/\sigma_r\simeq2.14$. This could be considered the actual embryo for a future more massive NSC;
\item for a nearly radial orbital decay, most of the cluster stellar mass is spread across the orbit and the resulting final cluster in the center of the galaxy is
poorly defined with a relatively low density ($\rho_{0}\simeq5\times10^{3}\,\mathrm{M_{\odot}/pc^3}$), an elongated shape and slow rotation compared to the average velocity dispersion ($v_r/\sigma_r\simeq0.33$). Such situation would likely prevent the formation of a real NSC;
\item for low eccentricity orbits, the final mass distribution in the innermost region of the merger remnant shows an overabundance of stars from the massive K1 cluster ($M_{K1}/M_{e}=0.77$), while in the initially quasi-radial orbits the resulting central region population is much more mixed; these features would result in different expectation of the inner stellar population in terms of age and metallicity between the two scenarios;

 In spite of the high level of uncertainty on the initial conditions for the two clusters, our simulations provide constraints on the merger timescale, which is significantly short. Since our study is limited to a very small part of the phase-space, our results have to be considered as boundary conditions and an indicative guideline for further systematic studies. From observational point of view, a higher spatial and spectral resolution of such systems is required in order to be able to provide more robust empirical constraints to enable detailed modeling, including dynamical friction, which depends on the knowledge of local velocity field and density.

\end{enumerate}

\section{Acknowledgements}
This research has made use of the NASA/IPAC Extragalactic Database (NED), which is funded by the National Aeronautics and Space Administration and operated by the California Institute of Technology.\\
We acknowledge the usage of the HyperLeda database \citep{Makarov14}.

\section*{Data availability}
The data underlying this article will be shared on reasonable request to the corresponding author.

\bibliographystyle{mnras}
\bibliography{references}

\end{document}